\documentclass[a4paper,12pt]{article}

\title{
\begin{flushright}
{\small INR-TH-2014-011}
\end{flushright}
Investigation of Q-tubes stability using the piecewise parabolic potential}

\usepackage{color}
\usepackage[english]{babel}
\usepackage{amssymb,amsmath}
\usepackage[pdftex]{graphicx}
\usepackage[pdftex]{hyperref}
\usepackage[left=3cm,right=2.5cm,top=2cm,bottom=3cm,bindingoffset=0cm]{geometry}
\graphicspath{{}}
\begin{document}
\author{
 E.\;Nugaev$^a$\thanks{{\bf e-mail}:
emin@ms2.inr.ac.ru}, A.\;Shkerin$^{a,b}$\thanks{{\bf e-mail}:
shkerin@inr.ru}
\\
$^a${\small{\em
Institute for Nuclear Research of the Russian Academy
of Sciences,}}\\
{\small{\em 60th October Anniversary prospect 7a, 117312, Moscow,
Russia
}}\\
$^b${\small{\em Moscow Institute of Physics and Technology,
}}\\
{\small{\em  Institutskii per., 9, 141700, Dolgoprudny, Moscow Region, Russia}}}
\date{}
\maketitle
\begin{abstract}
We analyze the classical stability of Q-tubes --- charged extended objects in $(3+1)$-dimensional complex scalar field theory. 
Explicit solutions were found analytically in the piecewise parabolic potential.
Our choice of potential
allows us to construct a powerful method of stability investigation.
We check that in the case of the zero winding number $n=0$, 
the previously known stability condition $\partial^2E/\partial Q^2<0$ for Q-balls is fulfilled.
However, in the case $n\geq 1$, we find 
a continuous family of instabilities. 
Our result has an analogy with the theory of superconductivity of the second type,
in which the vortex with $n>1$ becomes unstable towards the decay into the
$n$ vortices with the single winding number.
\end{abstract}

\renewcommand\bibname{\Large References}

\section*{1 INTRODUCTION}

In the variety of nontopological solitons, tube-like defects with nontrivial
winding number $n$
were discovered
\cite{Volkov:2002aj} relatively not long ago. In the theory of the single complex
scalar field with self-interaction, their features are similar to those of Q-balls.
The latter type of solitons has an interesting criterion of classical stability,
i.e., $\partial^2E/\partial Q^2<0$, were $E$ is the soliton energy and $Q$ is the global
charge\footnote{For Q-balls, this criterion was introduced in Ref.\cite{PhysRevD.13.2739}. Remarkably, it was derived earlier for solitons of the
nonlinear Schrödinger Equation (NSE); see Refs.\cite{1973R&QE...16..783V,1968JETP...26..994Z}}.
In this paper, the issue of Q-tube classical stability will be revisited. The reason
for our study is the possibility of transitions between solutions with different $n$. These transitions \cite{landau1996statistical},\cite{Bogomolny:1976tp}
do occur for ordinary Abrikosov-Nielesen-Olesen vortices in some
regions of the parameters of the Abelian Higgs model, and it is this fact that determines the
type of superconductivity. To consider transitions with the change of the winding number,
we will consider modes which are independent on the third spatial coordinate $z$ along the tube, i.e., we restrict ourselves to excitations in $(2+1)$-dimensional theory.
We will present explicit Q-tube solutions in a piecewise parabolic potential
and the method for the investigation of the stability for them. 
An advantage of our choice of potential is the separation of the radial equations
of motion for real and imaginary parts of the scalar field excitations except for the
finite number of the matching points.
In this case,
one can drastically simplify consideration and thoroughly survey instabilities
in the wide range of parameters. We will also present modes which are
responsible for the instabilities of solutions with $n\geq 1$ even for the case $\partial^2E/\partial Q^2<0$. On the other hand, in the case $n=0$ we did not find new instabilities, and our results correspond to the criterion of stability for Q-balls.
\section*{2 CLASSICAL SOLUTIONS}
\subsection*{2.1 Action}

To describe Coleman-type nontopological solitonS, let us consider the four-dimensional complex scalar field $\Phi$ with the action
\begin{equation}\label{Action}
S=\int d^4x\left(\partial_{\mu}\Phi^*\partial^{\mu}\Phi-V\left(\Phi^*\Phi\right)\right).
\end{equation}
To specify the axial symmetric Q-tube solution \cite{Volkov:2002aj} (see also
Ref.\cite{Radu:2008pp}), we will use the cylindrical coordinates $(r,\phi,z)$. 
Then the time-depended ansatz for the scalar field is
\begin{equation}\label{Q-tube profile}
\Phi(r,\phi,t)=F(r)e^{i\omega t}e^{in\phi}.
\end{equation}
Here $\omega$ is the continuous parameter of the solution, $n$ is an integer parameter, and $F(r)$ is some smooth real function.
To obtain a solution localized near the $z$ axis finite linear energy 
density, one should impose the following boundary 
conditions on $F(r)$ \cite{Volkov:2002aj}:
\begin{equation}\label{F-asymptotics}
\begin{array}{c}
F(0)=F^{(1)}(0)=...=F^{(n-1)}(0)=0, \\
F(r) \to 0, \qquad r\to\infty .
\end{array}
\end{equation}
The nonvanishing $tt$ component of the energy-momentum tensor for ansatz (\ref{Q-tube profile}) determines the energy:
\begin{equation}\label{E and J}
\begin{array}{ll}
E=\int d^3xT_{tt}, & T_{tt}=\left(\omega^2+\dfrac{n^2}{r^2}\right)F^2+\left(\dfrac{dF}{dr}\right)^2+V\left(F^2\right).
\end{array}
\end{equation}
There also exists a nonzero angular momentum $J=\int d^3xM^0_{xy}$, where $M^{\mu}_{\nu\rho}$ is the conserved current associated with the Lorentz invariance of the action in Eq.(\ref{Action}), and $(x,y)$ are Cartesian coordinates $x=r\cos\phi$, $y=r\sin\phi$. Using the relation $M^{\mu}_{\nu\rho}=T^{\mu}_{\nu}x_{\rho}-T^{\mu}_{\rho}x_{\nu}$, which is valid for a spinless field $\Phi$, one can reduce the expression for $J$ to the form
\begin{equation}\label{J}
J=2n\omega\int d^3x F^2.
\end{equation}
In addition, the internal $U(1)$ symmetry of the action implies the conservation of charge
\begin{equation}\label{Q}
Q=-i\int d^3x\left(\Phi^*\dot{\Phi}-\dot{\Phi^*}\Phi\right)=2\omega\int d^3x F^2=J/n.
\end{equation}
This connection between charge and angular momentum was found before in the theory of rotating boson stars \cite{Schunck:1996he}.
As will be seen later, the integer parameter $n$ separates Q-tube configurations into discrete domains. To find a classical soliton solution with a given $J$ and $Q$, one can search for an extremum of the functional
\begin{equation*}
E-\left(\lambda_1+n\lambda_2\right)\left(2\omega\int d^3xF^2-Q\right),
\end{equation*}
where $\lambda_1$, $\lambda_2$ are Lagrange multipliers. Taking the derivative with respect to $\omega$, one can obtain $\lambda_1+n\lambda_2=\omega$. Variation on $F$ leads to
\begin{equation}\label{F-equation}
\dfrac{d^2F}{dr^2}+\dfrac{1}{r}\dfrac{dF}{dr}-\dfrac{n^2}{r^2}F+\omega^2F=\dfrac{dV}{dF^2}F.
\end{equation}
This result certainly coincides with Lagrangian equation of motion derived from
action (\ref{Action}) using ansatz (\ref{Q-tube profile}). It should be noted that, due to $z$ independence of the tubes, the quantities calculated by Eqs.(\ref{E and J})-(\ref{Q}) all diverge. To avoid the confusion, from now on, when speaking about (3+1)-dimensional tubes, we will consider their energies, charges, and angular momenta only per unit length, but we will still denote them by the same letters $E$, $Q$, and $J$.

\subsection*{2.2 Potential and solutions}
As has been shown in Refs.\cite{Rosen,PhysRevD.61.047701}, the explicit analytic solution for Q-balls can be obtained by setting the potential $V$ to have a piecewise parabolic form. 
In the case of Q-tubes, we will use the following form of $V$
for the action (\ref{Action}):
\begin{equation}\label{Potential}
V\left(\vert\Phi\vert^2\right)=M^2\vert\Phi\vert^2\theta\left(1-\dfrac{\vert\Phi\vert^2}{v^2}\right)+\left(m^2\vert\Phi\vert^2+\Lambda\right)\theta\left(\dfrac{\vert\Phi\vert^2}{v^2}-1\right),
\end{equation}
which turns out to be more convenient for the investigation of 
stability \cite{Gulamov:2013ema}. Here $M^2>0$, $v>0$ and $m^2<M^2$ are parameters of the model\footnote{Values $m^2<0$ are also in consideration. In this case, one can add positive terms to the potential for large values of the field modulus without affecting the physics at the scale we are interested in.}; $\theta$ is the Heaviside step function with the convention $\theta(0)=\frac{1}{2}$; and $\Lambda=v^2(M^2-m^2)$ provides continuity of potential at the point $\vert\Phi\vert^2=v^2$. With this potential, Eq.(\ref{F-equation}) becomes
\begin{equation}\label{Q-tube Equation}
r^2\dfrac{d^2F}{dr^2}+r\dfrac{dF}{dr}+F\left(r^2\left(\omega^2-U\right)-n^2\right)=0,
\end{equation}
where $U=M^2$ for $F^2<v^2$ and $U=m^2$ otherwise. Equation (\ref{Q-tube Equation}) can be reduced to the Bessel equation for all $r$ except the matching points $r=r_i$, $i=1,..,N$, at which $F^2=v^2$. One can solve it separately in the intervals $(0,r_1)$, $(r_{i-1},r_i)$, and $(r_N,\infty)$. Due to the linearity, its general solution in each interval of $r$ contains two independent functions multiplied by arbitrary constants $C_i$. Boundary conditions and the smoothness requirement determine the particular solution from a general one in each interval of $r$. To obtain Q-tube solutions, it is sufficient to consider the case $N=1,2$.

Let us consider the case $N=2$. The finiteness of Q-tube energy per unit length demands regularity at infinity. We can obtain a required suppression of $F$ at $r\rightarrow\infty$ only if we impose $\omega^2-M^2<0$ for $U=M^2$ (i.e., for $r>r_2$ and $r<r_1$). The appropriate solution of Eq.(\ref{Q-tube Equation}) at $r>r_2$ is the MacDonald function $C_4K_n(\sqrt{M^2-\omega^2}r)$ with some constant $C_4$. In the region $r<r_1$, the only applicable solution for $n>0$ is the Infeld function, $C_1I_n(\sqrt{M^2-\omega^2}r)$. 
It should be mentioned that both type of solutions are monotonic functions.
To construct $F$ between $r_1$ and $r_2$, where $U=m^2$, one should suppose $\omega^2-m^2>0$ and take the linear combination of Bessel functions of the first and the second kinds: $C_2J_n(\sqrt{\omega^2-m^2}r)+C_3Y_n(\sqrt{\omega^2-m^2}r)$.
This choice allows to obtain solutions which are equal to the same value
at $r_1$ and $r_2$ due to the oscillating character of the corresponding equation.
Choosing constants $C_i,i=1,..,4$ appropriately, one can obtain the smooth Q-tube profile with a given $n>0$ and $m^2<\omega^2<M^2$. 

\begin{figure}[h!]
\begin{minipage}[h]{0.49\linewidth}
\center{\includegraphics[width=1.0\linewidth]{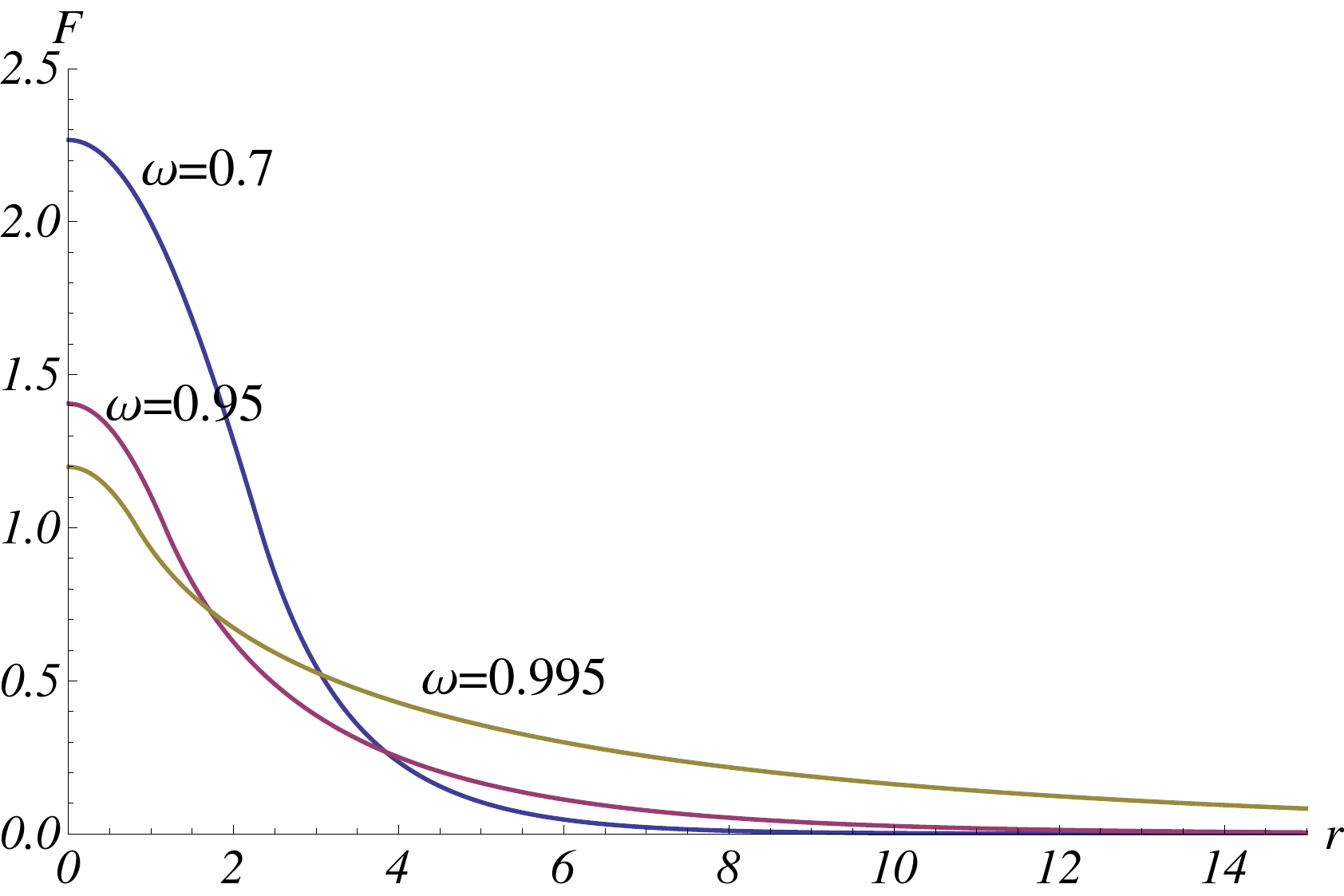}}
\end{minipage}
\begin{minipage}[h]{0.49\linewidth}
\center{\includegraphics[width=1.0\linewidth]{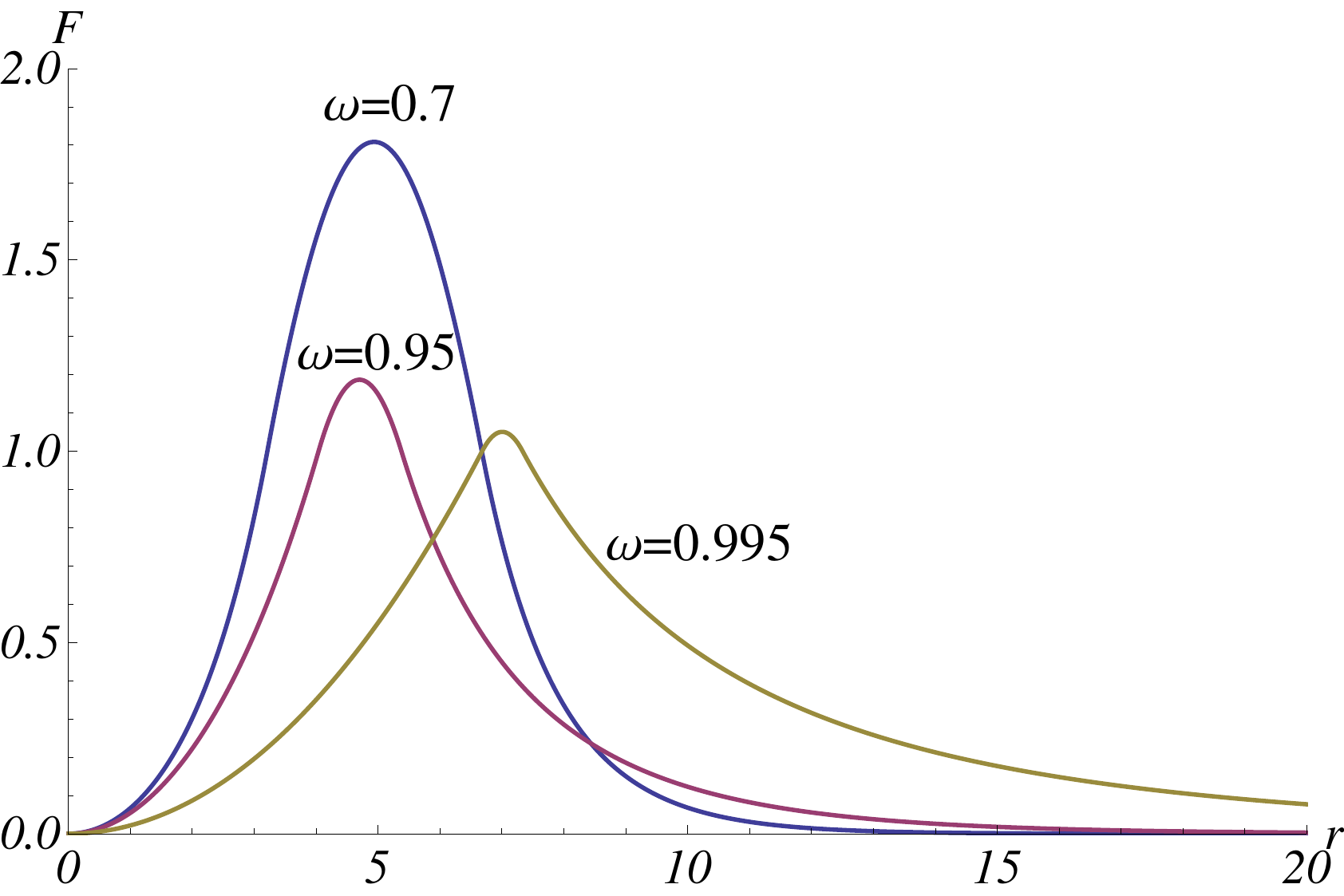}}
\end{minipage}
\caption{Q-tube profiles for different values of $\omega$ with $m=0$, $M=v=1$, and $n=0$ (left plot), $n=2$ (right plot).} 
\label{Q-tube profiles}
\end{figure}

The configuration with $n=0$ is similar to a usual Q-ball in three
spatial dimensions which is described by a monotonically decreasing function of radius.
One can easily obtain them in a sector with $N=1$, corresponding to the single 
point, $r=r_0$, at which the different solutions of Eq.(\ref{Q-tube Equation}) should be matched. We take again the MacDonald function $K_0(\sqrt{M^2-\omega^2}r)$ at $r>r_0$ and smoothly match it to $J_0(\sqrt{\omega^2-m^2}r)$, which presents $F$ at $r<r_0$. In Fig.\ref{Q-tube profiles}, we present several examples of profiles. It should be noted that the behaviour of $K_n$ and $I_n$ at infinity and the origin, correspondingly, provide the required asymptotics [Eq.(\ref{F-asymptotics})] for $F$. 

\subsection*{2.3 Properties of the solutions}
The form of $F$ depends on the particular choice of $\omega$. One can formally denoted by $F(r;\omega)$ the continuous family of Q-tube configurations corresponding to the specific values of $\omega$. By substituting it into the integrands in Eqs.(\ref{E and J}) and (\ref{Q}) and taking the derivative with respect to $\omega$, it is possible to obtain the relation
\begin{equation*}
\dfrac{dE}{d\omega}=\omega\dfrac{dQ}{d\omega},
\end{equation*}
which holds for any Q-ball solution ~\cite{PhysRevD.13.2739,Lee:1991ax,tsumagari2009physics}. This relation indicates that a parametric plot $E(Q)$ may contains cusps at some values $\omega_c$, corresponding to the simultaneous extrema of $E$ and $Q$. We also use
this relation to check our numerical calculations.

\begin{figure}[h!]
\begin{minipage}[h]{0.49\linewidth}
\center{\includegraphics[width=1.0\linewidth]{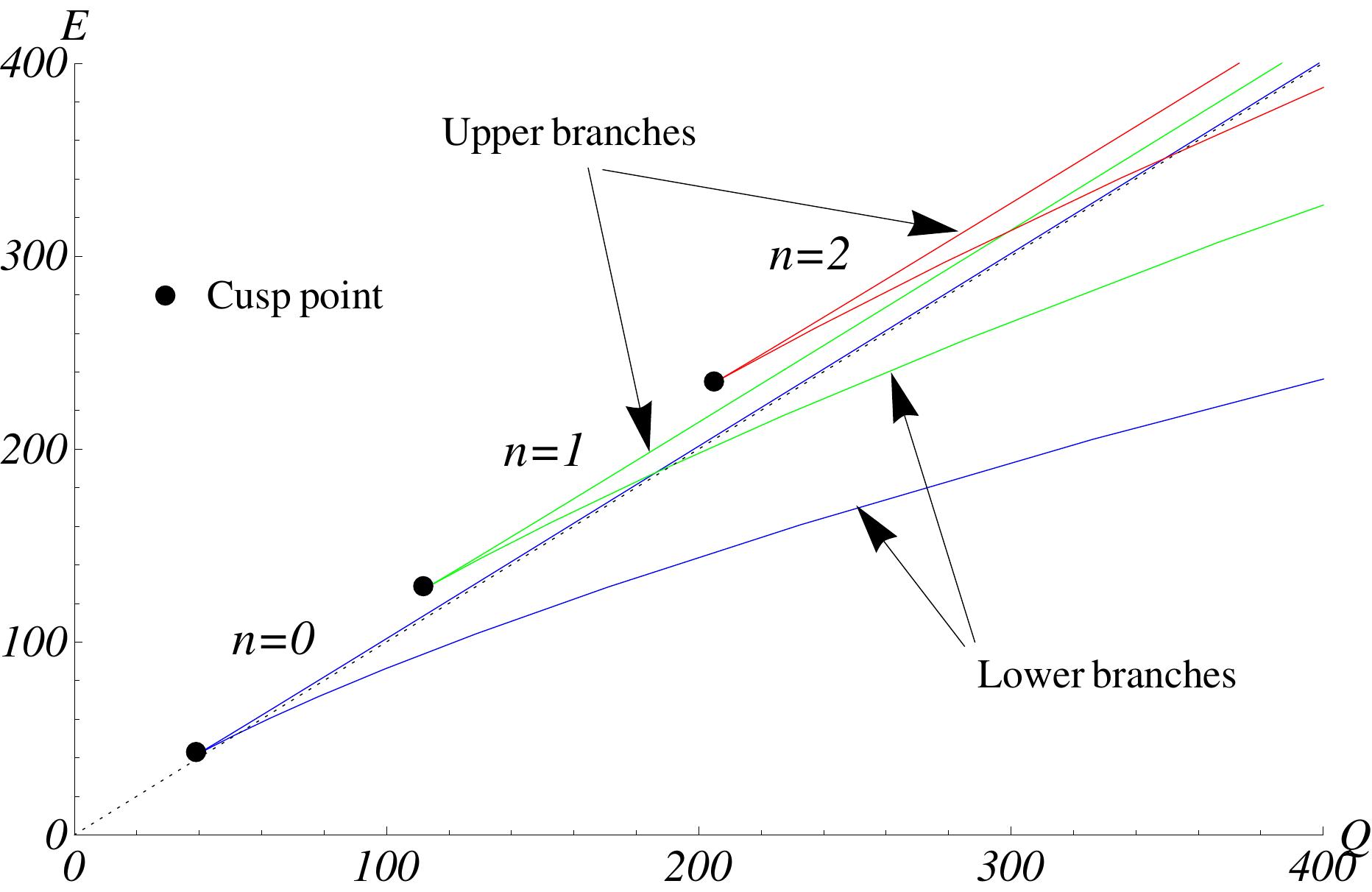}\\a}
\label{E(Q)positive m}
\end{minipage}
\begin{minipage}[h]{0.49\linewidth}
\center{\includegraphics[width=1.0\linewidth]{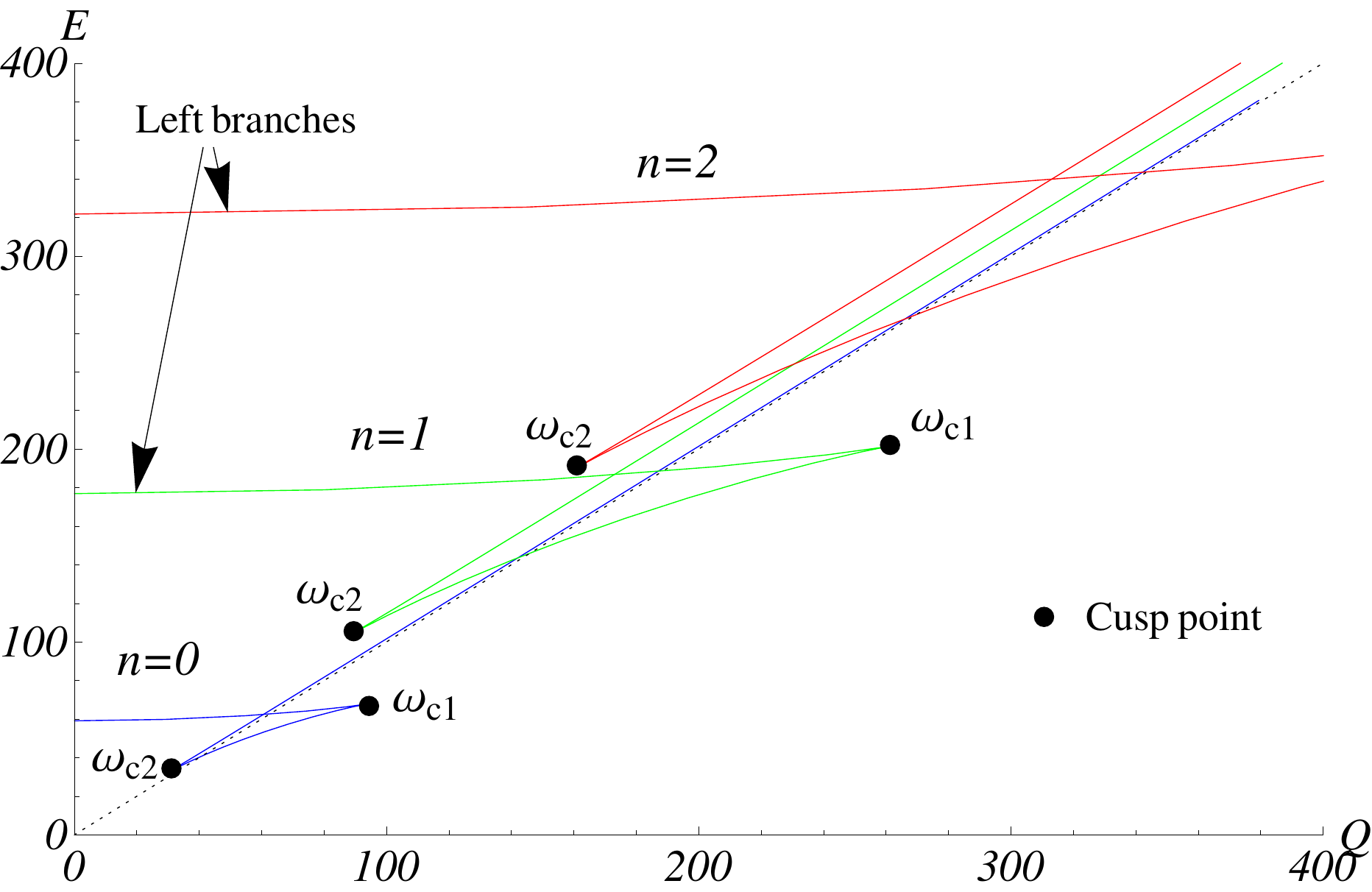}\\b}
\label{E(Q)negative m}
\end{minipage}
\caption{$E(Q)$ for the unit-length Q-tubes for different values of $n$ with $M=v=1$ and $m^2\geqslant 0$ (left plot) and $m^2<0$ (right plot). The dotted line through the
origin $E=MQ$ corresponds to the free particles at rest.} 
\label{E(Q)}
\end{figure}

In Fig.\ref{E(Q)}a, we present the $E(Q)$ dependence, with $\omega$ being the parameter and $m^2\geqslant 0$. The asymptote of the lower branch corresponds to the limit $\omega\rightarrow m$, and the upper branch corresponds to $\omega\rightarrow M$. The solution at the cusp $\omega=\omega_c$ determines the minimal energy density $E_{min}$ and charge density $Q_{min}$. For every density $Q>Q_{min}$ there are two Q-tube configurations. Living on the lower branch, Q-tubes have quite localized profiles: their energy density is concentrated at $r_1<r<r_2$. The upper branch corresponds to the less localized objects. The energy integral [Eq.(\ref{E and J})] for them is determined by the tail, at $r>r_2$, despite the exponential suppression of $F$ in this interval, provided by the MacDonald function; see Fig.\ref{Q-tube profiles}. 

The case $m^2<0$ provides the different picture shown in Fig.\ref{E(Q)}b. There is  a nontrivial solution with zero $Q$ that occurs at $\omega=0$, as follows from (\ref{Q}). The dependence $E(Q)$ contains two cusps, at $\omega_{c1}$ and $\omega_{c2}$.

These $E(Q)$ dependences for Q-tubes of unit length are similar to those for Q-balls. For instance, the explicit four- and two-dimensional Q-ball solutions with potential of the form given in Eq.(\ref{Potential}), studied in Ref.\cite{Gulamov:2013ema}, have the same $E(Q)$ behaviour.

\section*{3. CLASSICAL STABILITY}

\subsection*{3.1 Preliminaries}
Figure \ref{E(Q)} demonstrates that while $n$ increases, the energy of a Q-tube with a given $Q$ also increases. One can then suppose the existence of transitions between the branches with different $n$, as far as transitions between upper and lower branches with the same $n$, ruled by $E$, $Q$, and $J$ conservation laws. They would correspond to the decay of the initial Q-tube into one or several Q-tubes with their own $n$'s as such a multitube state may turn out to be more energetically favourable in analogy with Abelian vortices \cite{landau1996statistical},\cite{Bogomolny:1976tp}. We will try to search for these transitions below and restrict ourselves to the investigation of the small perturbations without dependence on the spatial coordinate $z$. Equivalently, one can say that we study (2+1)-dimensional solutions, with $E$, $Q$, and $J$ replaced by their analogues in (2+1) dimensions. Thus, the possibility of a Q-tube's division into pieces at some breaking point $z=z_0$ is outside of our attention. The criterion of the
applicability of the classical consideration will be discussed in Sec. 3.5.

\subsection*{3.2 Solution of linearized equation of motion}
Let us return to the original action [Eq.(\ref{Action})]. The variational principle leads to the equation for $\Phi$ in cylindrical coordinates $(r,\phi,z)$ (we assume that $\Phi$ does not depend on $z$),
\begin{equation}\label{f-equation}
r^2\dfrac{d^2\Phi}{dr^2}+r\dfrac{d\Phi}{dr}+\dfrac{d^2\Phi}{d\phi^2}-r^2\dfrac{d^2\Phi}{dt^2}=r^2\Phi\dfrac{dV(\Phi^*\Phi)},{d\vert\Phi\vert^2}
\end{equation}
and the complex conjugated equation on $\Phi^*$.
The general $U(1)$-invariant potential $V$ dependence on $\Phi^*$ in Eq.(\ref{f-equation}) leads to mixing between $\Phi$ and $\Phi^*$ for each point $r$.
Consider the solution of Eq.(\ref{f-equation}) of the form $\Phi=\Phi_0+h$, where $\Phi_0$ is the Q-tube solution [Eq.(\ref{Q-tube profile})] and $h=h(r,\phi,t)$ --- some small complex perturbation. 
We are interested in exponentially growing modes which indicate the existence of the classical instability. We would also like to separate variables in Eq.(\ref{f-equation}). For this, we take the ansatz first proposed in Ref.\cite{Anderson:1970et} and slightly modify it for the case of nonzero angular momentum $(n\pm l)$, $l\geqslant 0$:
\begin{equation}\label{anzats}
h=e^{i\omega t+in\phi}\sum_{l=0}^{\infty}\left(c_1^le^{i(\alpha t+l\phi)}+c_2^{l*}e^{-i(\alpha^* t+l\phi)}\right),
\end{equation}
where $c_{1,2}^l$ are some functions of $r$. By substituting  $\Phi$ into Eq.(\ref{f-equation}) and extracting the terms with equal exponential factors, 
one can obtain following equations for $c_{1,2}^l(r)$ (we omit the index $l$ below):
\begin{equation}\label{Pert. Equations}
\begin{array}{r}
r^2\dfrac{d^2c_1}{dr^2}+r\dfrac{dc_1}{dr}-c_1\left[r^2\left(\left(\gamma +i(\omega +\gamma ')\right)^2+U\right)+(n+l)^2\right]=\\
=\dfrac{r^2F^2}{v^2}\left(m^2-M^2\right)\delta\left(\dfrac{F^2}{v^2}-1\right)(c_1+c_2),\\
r^2\dfrac{d^2c_2}{dr^2}+r\dfrac{dc_2}{dr}-c_2\left[r^2\left(\left(\gamma -i(\omega -\gamma ')\right)^2+U\right)+(n-l)^2\right]=\\
=\dfrac{r^2F^2}{v^2}\left(m^2-M^2\right)\delta\left(\dfrac{F^2}{v^2}-1\right)(c_1+c_2),
\end{array}
\end{equation}
where we set 
\[
\alpha=-i\gamma+\gamma ', \qquad\gamma,\gamma '\in\mathbb{R}.
\]
Except for the matching points $r=r_i$, the equations for $c_1$ and $c_2$ are separated. This crucial result determines our choice of the piecewise parabolic potential of the type given in Eq.(\ref{Potential}).  

In the next two sections, we will search for solutions of Eq.\ref{Pert. Equations} with $\gamma '=0$, i.e., we hold $\alpha$ to be purely imaginary. This simple choice is motivated by results of Ref.\cite{Gulamov:2013ema}, where all instabilities of Q-balls were found for $\gamma'=0$ only. The generalization on arbitrary $\alpha$ will be considered in Sec.3.5. Note that the modes with $l=0$ should be considered individually, since the exponential terms cannot help to separate $c_1$ and $c_2$. We suppose that the general consideration is applicable to this case for the appropriate limiting procedure $\gamma '\to 0$.

Equation (\ref{Pert. Equations}) gives Bessel equations except for the points $r=r_i$. Thus, the following discussion lies close to the case of Q-tubes. For $n>0$, there are three intervals ("left", $r<r_1$; "middle", $r_1<r<r_2$; and "right", $r>r_2$). In each interval, the general solution of Eq.(\ref{Pert. Equations}) contains two arbitrary constants. The requirement of the regularity at zero and infinity, following from the finiteness of the energy, fixes some of these constants, which can be turned to zero by the appropriate choice of the basis of solution. The other constants are needed for the matching of solutions of Eq.(\ref{Pert. Equations}) at the points $r=r_i$. Due to the delta functions, the modes we construct are continuous but not smooth.

\subsection*{3.3 The case $n=0,\gamma'=0$}

Let us illustrate this procedure by the simple example of a Q-tube with $n=0$. In this case, there is only one matching point. Examples for the background are
presented in the left plot of Fig.(\ref{Q-tube profiles}). Imposing boundary conditions at the origin and infinity, we obtain four arbitrary constants --- say, $C_i,i=1,..,4$. The continuity at the point $r=r_0$ forms two equations for these constants. Two more conditions follow from the integration of the delta functions in Eq.(\ref{Pert. Equations}). Writing out Eq.(\ref{Pert. Equations}) at this point, we find
\begin{equation}\label{system}
\begin{array}{c}
C_1c_{1,left}(r_0)-C_2c_{1,right}(r_0)=0, \\
C_3c_{2,left}(r_0)-C_4c_{2,right}(r_0)=0,\\
C_2c_{1,right}'(r_0)-C_1c_{1,left}'(r_0)-A\left(C_1c_{1,left}(r_0)+C_3c_{2,left}(r_0)\right)=0,\\
C_4c_{2,right}'(r_0)-C_3c_{2,left}'(r_0)-A\left(C_1c_{1,left}(r_0)+C_3c_{2,left}(r_0)\right)=0,\\
A=\dfrac{v\left(m^2-M^2\right)}{2\vert F'(r_0)\vert}.
\end{array}
\end{equation}
Thus, for a given Q-tube background, the growing mode with fixed $l$ exists when the determinant $\Delta$ of the system [Eq.(\ref{system})] vanishes, Re$\Delta(\gamma)=$Im$\Delta(\gamma)=0$, for some value of $\gamma>0$. For $n>0$, the number of conditions is doubled as much as the number of arbitrary constants.
We also used obvious zero modes due to the translational and internal symmetry
for our check of conditions in Eq.(\ref{system}).

Q-tubes with zero angular momentum require particular attention, since formally Eq.(\ref{F-equation}) for them coincides with that for a two-dimensional Q-ball. Therefore, one can use the analogy between those Q-tubes and various Q-ball solutions (e.g., studied in detail in Ref.\cite{Gulamov:2013ema}).

In accordance with the discussion above, one should find the basis solutions of Eq.(\ref{Pert. Equations}), $c_{1,2,left}$ and $c_{1,2,right}$, which would provide regularity at zero and infinity. Such functions are as follows:
\begin{equation}\label{n=0 sol}
\begin{array}{c}
c_{1,2,left}=J_{\pm l}\left(r\sqrt{\left(\omega\mp i\gamma \right)^2-m^2}\right),\\
c_{1,2,right}=H^{(1)}_{\pm l}\left(r\sqrt{\left(\omega\mp i\gamma \right)^2-M^2}\right),
\end{array}
\end{equation}
where $H^{(1)}_{l}$ is the Hankel function of the first kind.

Let us study the asymptote of $\Delta(\gamma )$, at which the solution in Eq.(\ref{n=0 sol}) turns to the more clear form. Namely, consider the case of large $\gamma $, $\gamma \gg \vert m\vert,M$. The arguments of the functions in Eq.(\ref{n=0 sol}) can be expanded in a series on $\gamma $. Then, holding the leading term on $\gamma $ and using the Bessel function addition theorem \cite{gradshteyn2007}, we have
\begin{equation}
\begin{array}{c}
c_{1,2,left}=e^{\pm\frac{i\pi l}{2}}I_l(r\gamma )+O(\sqrt{\gamma }),\\
c_{1,2,right}=\dfrac{2i}{\pi}e^{\mp\frac{i\pi l}{2}}K_l(r\gamma )+O(\sqrt{\gamma }).
\end{array}
\end{equation}
The complex factors before $I_l$ and $K_l$ can be absorbed by the  constants $C_i$, $i=1,..,4$. Using the asymptotes of $I_l$ and $K_l$ at infinity \cite{gradshteyn2007}, we find for $\Delta$ [Fig.\ref{Det,n=0}a]
\begin{equation}\label{Delta}
\begin{array}{cc}
Re\Delta(\gamma)=\dfrac{1}{r_0^2}\left(\dfrac{\vert A\vert}{\gamma}-1\right), & Im\Delta(\gamma)=0.
\end{array}
\end{equation}
It follows from the behaviour of the Q-tube solution that $\vert A\vert\rightarrow\infty$ if (and only if) $\omega\rightarrow M$. So, one can expect the existence of the growing mode with $\gamma \approx\vert A\vert$ on the upper branch of the $E(Q)$ plot and, in fact, the catastrophic instability of Q-tubes with $\omega\approx M$. We used Eq.(\ref{Delta}), which is independent of $l$, for the checking of our results; see Fig.\ref{Det,n=0}a.

In the opposite limit $\gamma\rightarrow 0$, there are obvious solutions of Eq.(\ref{Pert. Equations}) at $\gamma=0$. One of them, with $l=0$,  have the form $h\sim i\Phi_0$ and corresponds to the $U(1)$ global symmetry of action (\ref{Action}). Two more modes appear as a result of the breaking of the translational invariance along the axes $x$ and $y$ by the Q-tube configuration, $h\sim \Phi'_{0r}$, and characterized by $l=1$. Thus, for modes with $l=0,1$, we have $\Delta(0)=\Delta'(0)=0$. 
Our numerical calculations satisfy these statements. Substitution of the solution in Eq.(\ref{n=0 sol}) $\Delta(\gamma)$ results in the dependences shown in Figs.\ref{Det,n=0}b,c,d. It should be mentioned that in the case $\gamma'=0$, Im$\Delta(\gamma)=0$ for all $\gamma$ and $l$. In a sector with $l=0$, we see from Figs.\ref{Det,n=0}b,c that the root of the system [Eq.(\ref{system})] exists if the sign of Re$\Delta''(0)>0$. Calculations show that its changing occurs exactly at the cusps of $E(Q)$. Thus, the lower branch of $E(Q)$ plot is stable against the perturbations with $l=0$, and the upper and left (if any) ones are unstable (see Fig.\ref{E(Q)}). This conclusion agrees with the results of Refs.\cite{Gulamov:2013ema,Alford:1987vs}. In addition, Fig.\ref{Det,n=0}d shows the absence of the growing modes in a sector with $l=1$ and, in fact, with all $l>0$
for $\gamma'=0$. 

\begin{figure}[h!]
\begin{minipage}[h]{0.5\linewidth}
\center{\includegraphics[width=1.0\linewidth]{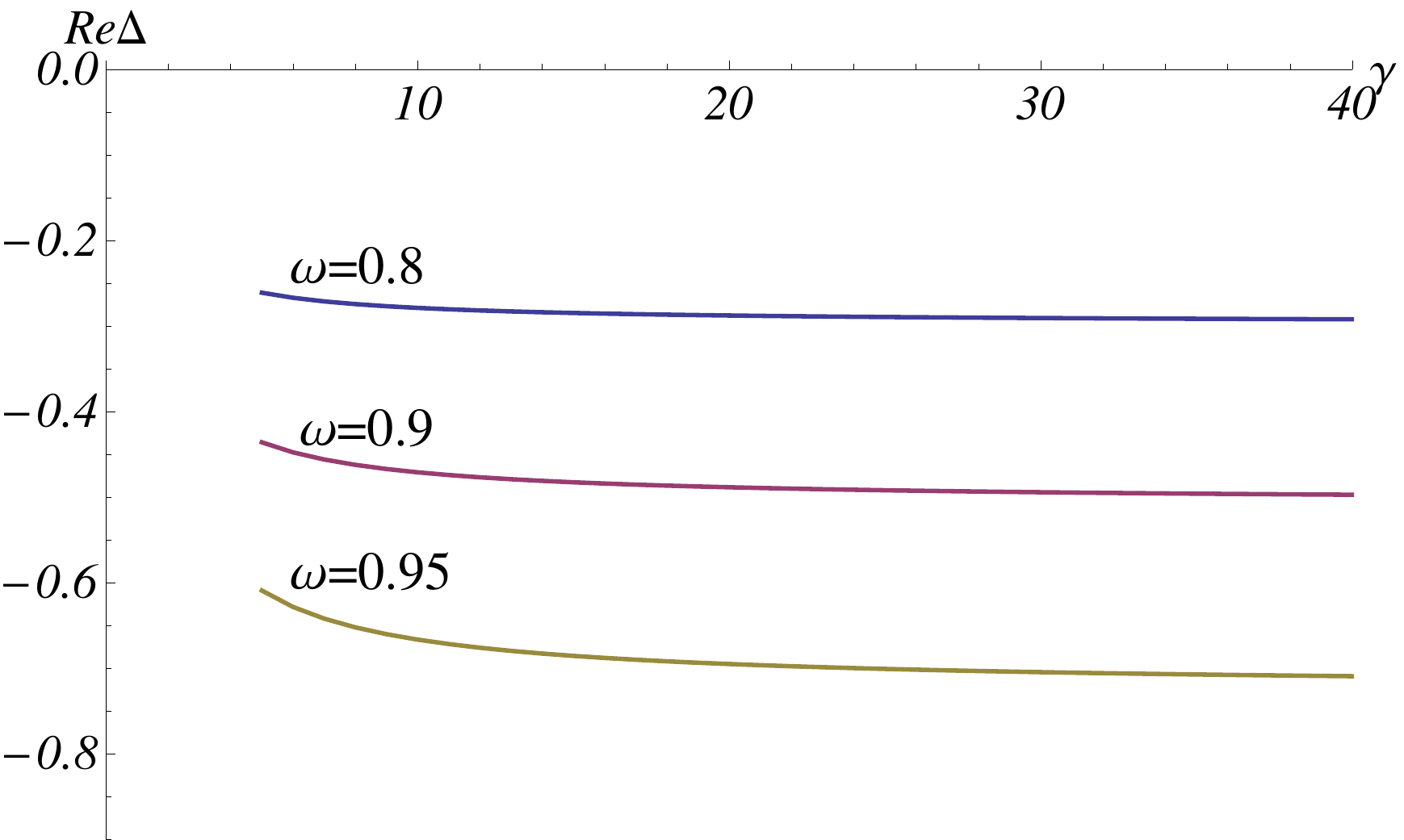}\\a}
\label{Large gamma,n=0}
\end{minipage}
\begin{minipage}[h]{0.5\linewidth}
\center{\includegraphics[width=1.0\linewidth]{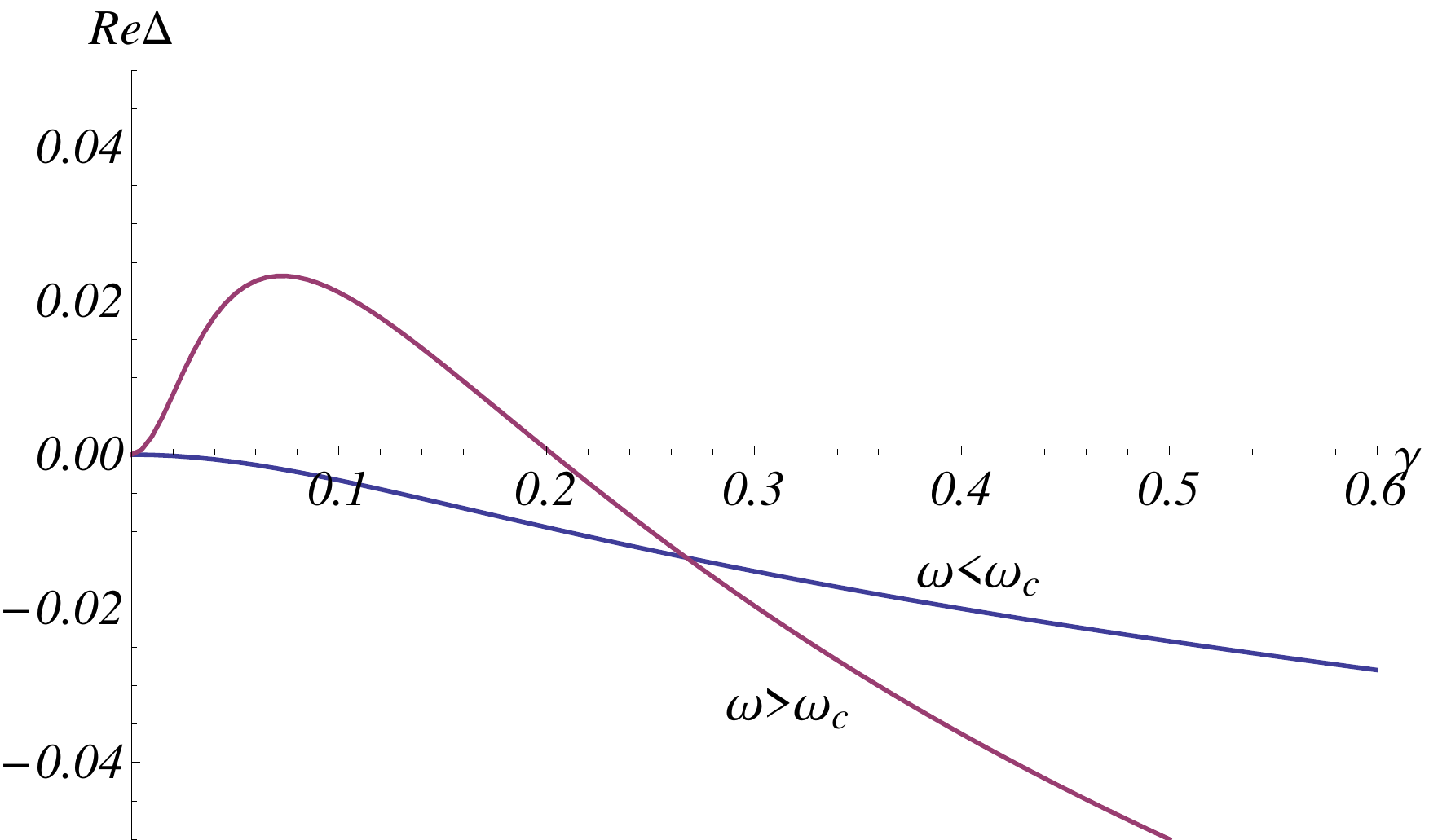}\\b}
\end{minipage}
\vfill
\begin{minipage}[h]{0.5\linewidth}
\center{\includegraphics[width=1.0\linewidth]{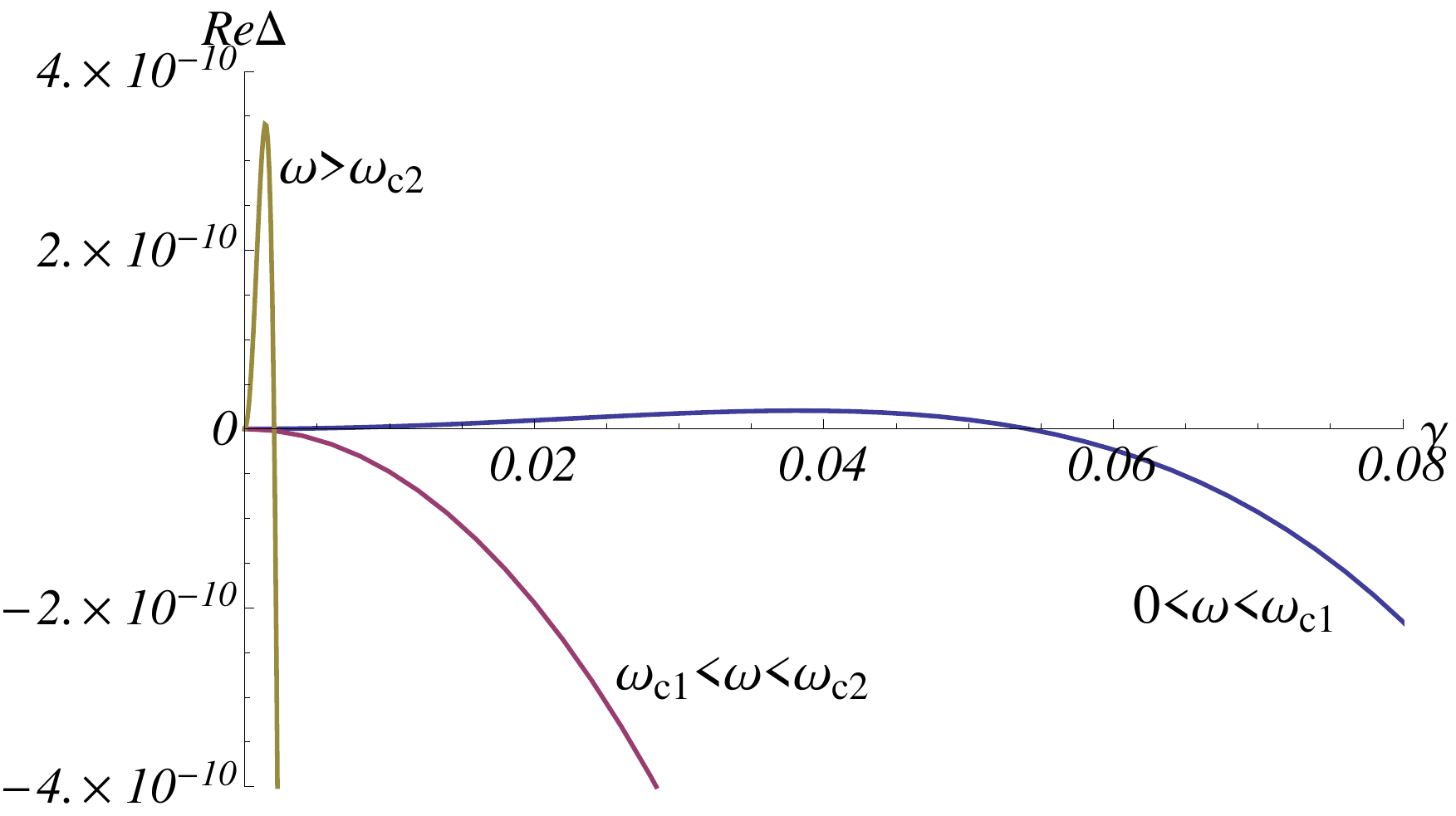}\\c}
\end{minipage}
\begin{minipage}[h]{0.5\linewidth}
\center{\includegraphics[width=1.0\linewidth]{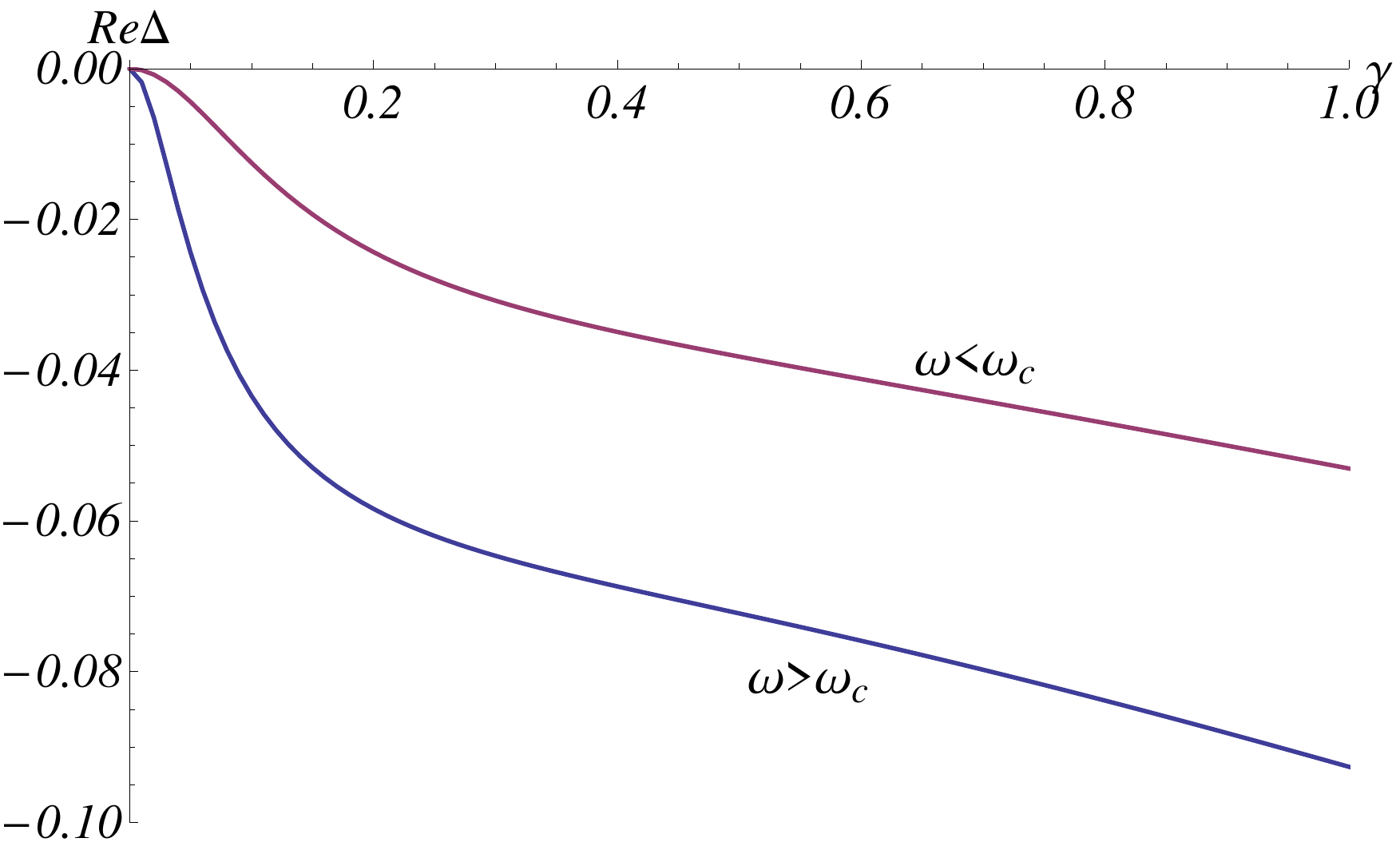}\\d}
\label{n=0,l=1}
\end{minipage}
\caption{$\Delta(\gamma)$ plots for $n=0$. a) The unified behaviour for large $\gamma$, shown is the example with $m=0$, $M=v=1$. b) Behaviour for $m^2\geqslant 0$, $l=0$. Parameters of the example: $m=0.1$, $M=v=1$, $\omega=0.94$ and $\omega=0.92$ ($\omega_c\approx 0.93$). c) Behaviour for $m^2<0$, $l=0$. Parameters of the example: $m^2=-0.25$, $M=v=1$, $\omega=0.1,0.85,0.93$ ($\omega_{c1}\approx 0.18$, $\omega_{c2}\approx 0.89$). d) The unified behaviour for $l=1$. Parameters of the example: $m=0$, $M=v=1$, $\omega=0.85,0.95$.} 
\label{Det,n=0}
\end{figure}

\subsection*{3.4 The case $n>0,\gamma'=0$}

\begin{figure}[h!]
\begin{minipage}[h]{0.5\linewidth}
\center{\includegraphics[width=1.0\linewidth]{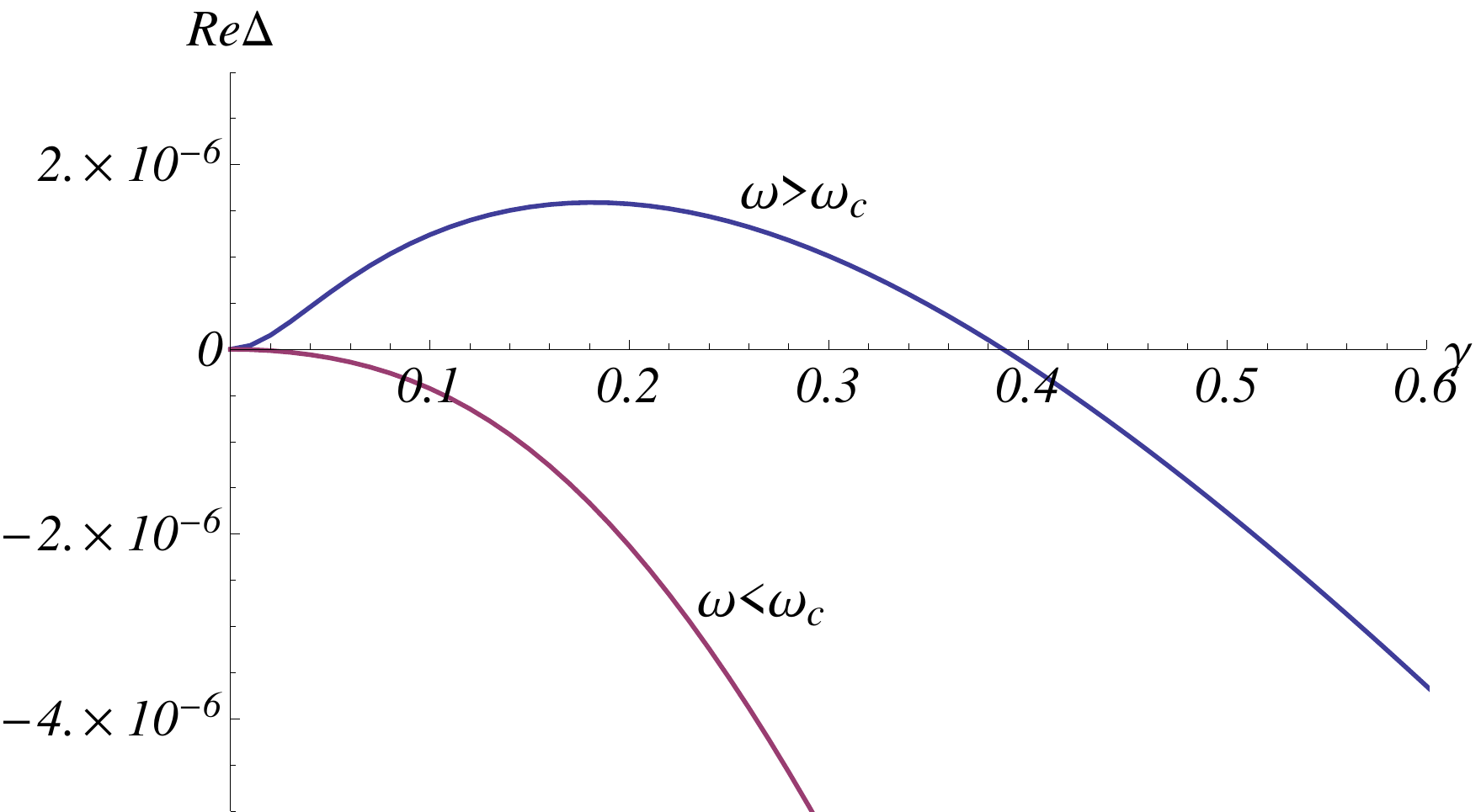}\\a}
\end{minipage}
\begin{minipage}[h]{0.5\linewidth}
\center{\includegraphics[width=1.0\linewidth]{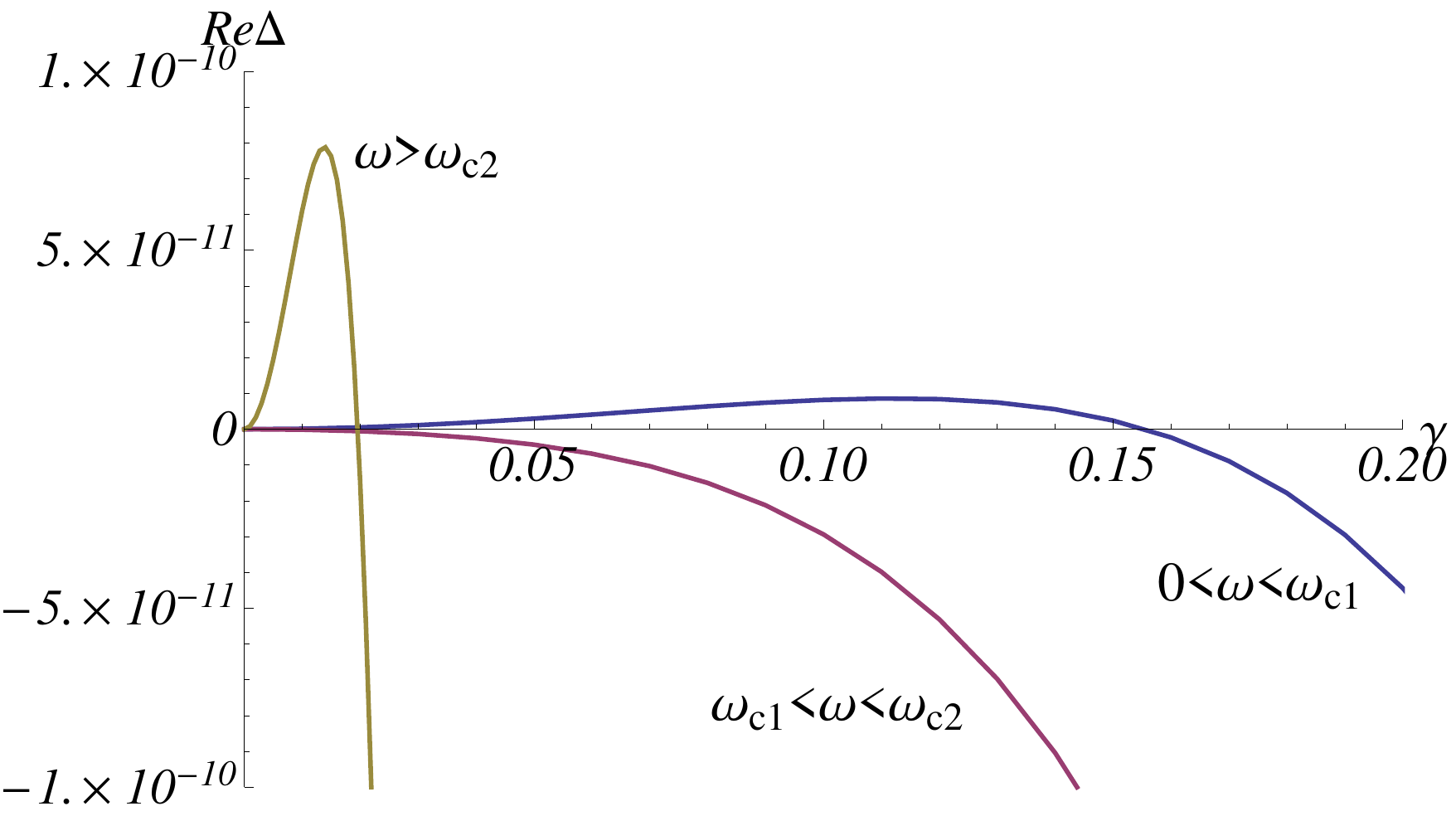}\\b}
\end{minipage}
\begin{minipage}[h]{0.5\linewidth}
\center{\includegraphics[width=1.0\linewidth]{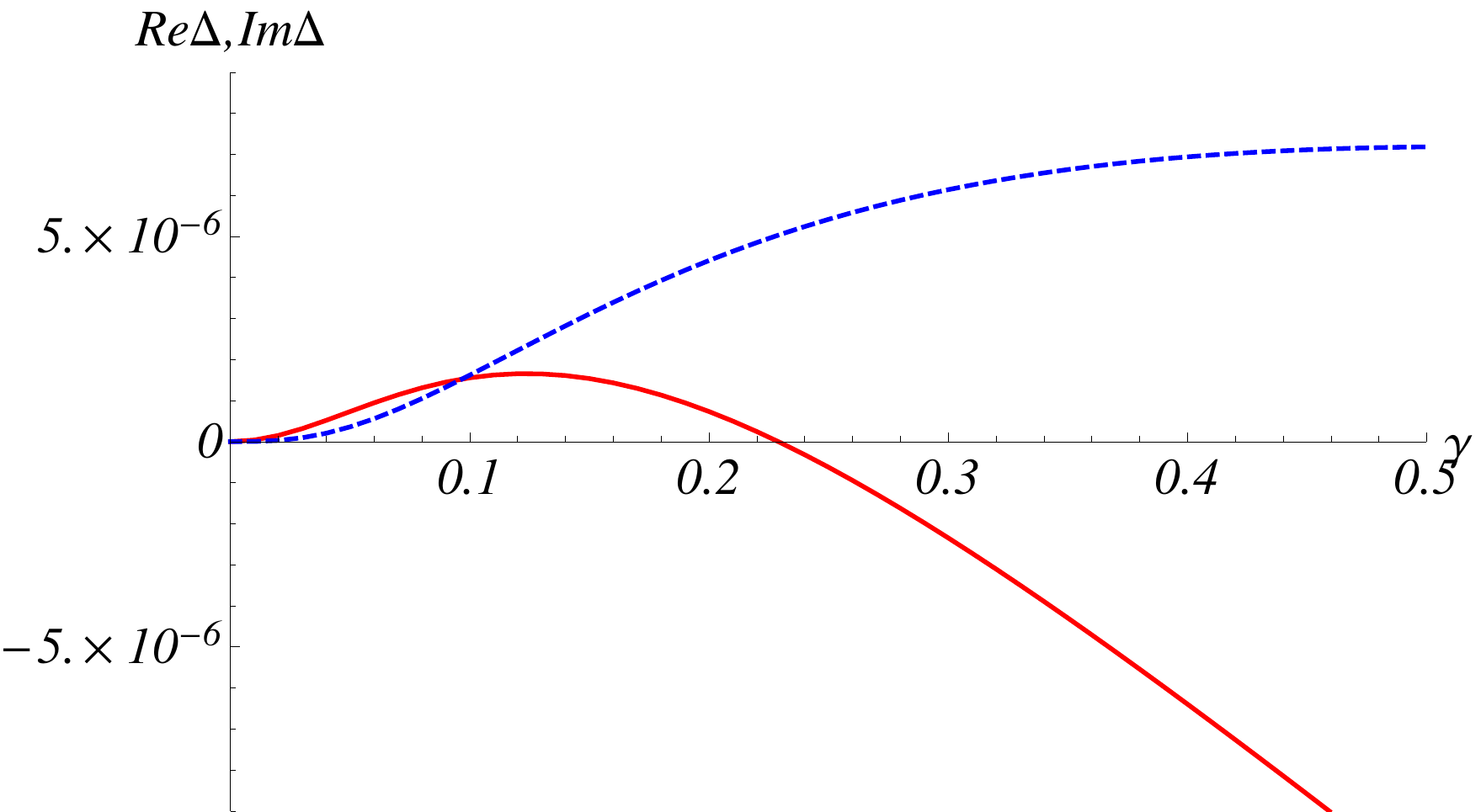}\\c}
\end{minipage}
\begin{minipage}[h]{0.5\linewidth}
\center{\includegraphics[width=1.0\linewidth]{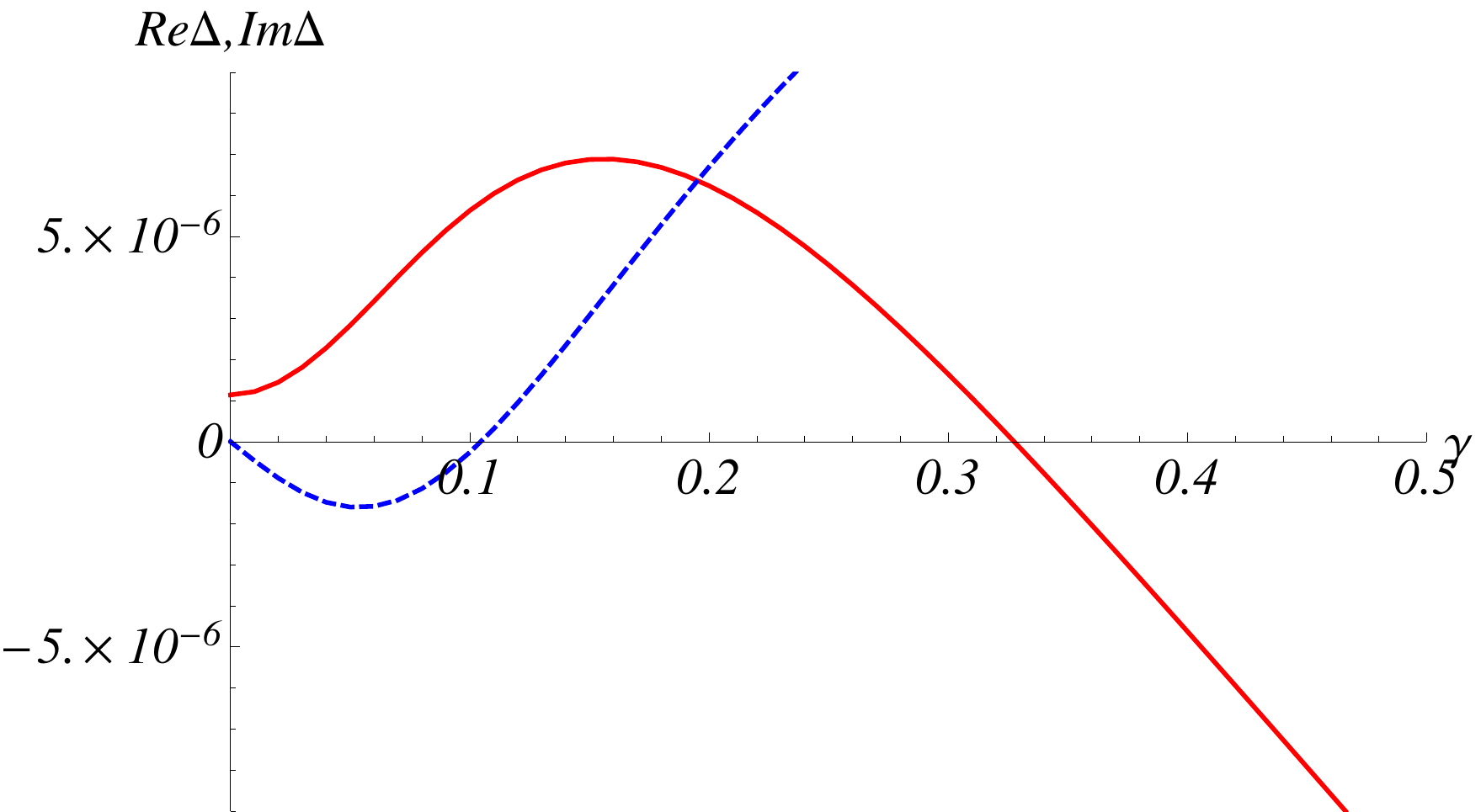}\\d}
\end{minipage}
\caption{$\Delta(\gamma)$ plots for $n>0$ ($n=2$ on the examples). a) Behaviour for $m^2\geqslant 0$, $l=0$. Parameters of the example: $m=0.1$, $M=v=1$, $\omega=0.98$, and $\omega=0.91$ ($\omega_c\approx 0.93$). b) Behaviour for $m^2<0$, $l=0$. Parameters of the example: $m^2=-0.25$, $M=v=1$, $\omega=0.1,0.85,0.93$ ($\omega_{c1}\approx 0.19$, $\omega_{c2}\approx 0.91$). c) General behaviour for perturbations with $l=1$. Shown are the real (solid line) and imaginary (dashed line) parts of $\Delta$. Parameters of the example: $m=0$, $M=v=1$, and $\omega=0.95$. d) General behaviour for perturbations with $l>1$. Parameters of the example: $m=0$, $M=v=1$, $\omega=0.75$, and $l=2$.}
\label{Det,n>0}
\end{figure}

The general consideration follows the particular case $n=0$ closely, except that the number of solutions that need to be matched is doubled. Since there is no regularity conditions imposed on the middle solution, $c_{1,2,middle}$, the basis of solutions at $r_1<r<r_2$ contains two independent functions. So, from Eq.(\ref{Pert. Equations}) we have
\begin{equation}\label{n>0 sol}
\begin{array}{l}
c_{1,2,left}=J_{n\pm l}\left(r\sqrt{\left(\omega\mp i\gamma \right)^2-M^2}\right),\\
c_{1,2,right}=H^{(1)}_{n\pm l}\left(r\sqrt{\left(\omega\mp i\gamma \right)^2-M^2}\right),\\
c_{1,2,middle 1}=J_{n\pm l}\left(r\sqrt{\left(\omega\mp i\gamma \right)^2-m^2}\right),\\
c_{1,2,middle 2}=H^{(1)}_{n\pm l}\left(r\sqrt{\left(\omega\mp i\gamma \right)^2-m^2}\right).
\end{array}
\end{equation}
By analogy with the case studied above, in the limit $\gamma \rightarrow\infty$ these functions turn to
\begin{equation}
\begin{array}{c}
c_{1,2,left}=c_{1,2,middle 1}=I_{n\pm l}(\gamma r),\\
c_{1,2,right}=c_{1,2,middle 2}=K_{n\pm l}(\gamma r)
\end{array}
\end{equation}
(We omit some constant multipliers), and the behaviour of $\Delta(\gamma)$ in this limit is identical to that for $n=0$. 

In the general case, substitution of Eq.(\ref{n>0 sol}) into the expression for $\Delta(\gamma)$ leads to the behaviour shown in Figs.\ref{Det,n>0}a,b. Again, for $l=0$, we have Im$\Delta(\gamma)=0$, and the sign of Re$\Delta''(0)$ correlates with the sign of $dQ/d\omega$, which indicates the instability of the upper and (if any) the left branches of $E(Q)$ plot. It also denotes the stability of the lower branch in the channel with $l=0$.

However, in a sector with $l>0$, the value of Im$\Delta(\gamma)$ becomes nonzero, and the equation $\Delta(\gamma)=0$ turns to two independent equations Re$\Delta(\gamma)=0$ and Im$\Delta(\gamma)=0$ for a single real variable $\gamma$. Calculation shows the absence of simultaneous roots of these equations for $\gamma'=0$, which indicates the absence of growing modes with $l>0$ at all $\omega$ and $m^2$.
In the case $\gamma'\neq 0$, however, the number of free parameters coincides with the number of
equations, and we will study this general case in the next section.

\begin{figure}[htb!]
\begin{minipage}[h]{0.5\linewidth}
\center{\includegraphics[width=1.0\linewidth]{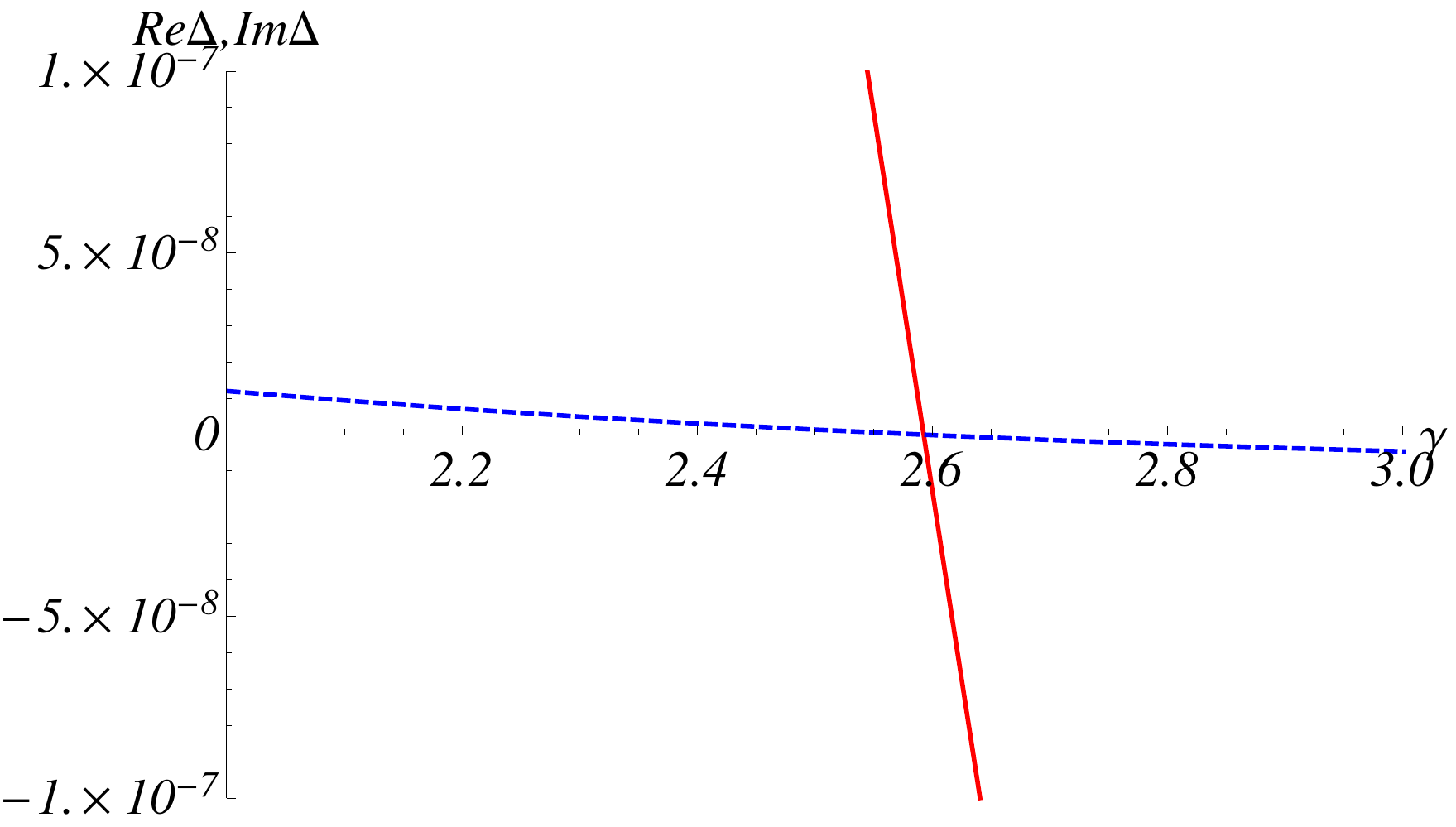}\\a}
\end{minipage}
\begin{minipage}[h]{0.5\linewidth}
\center{\includegraphics[width=1.0\linewidth]{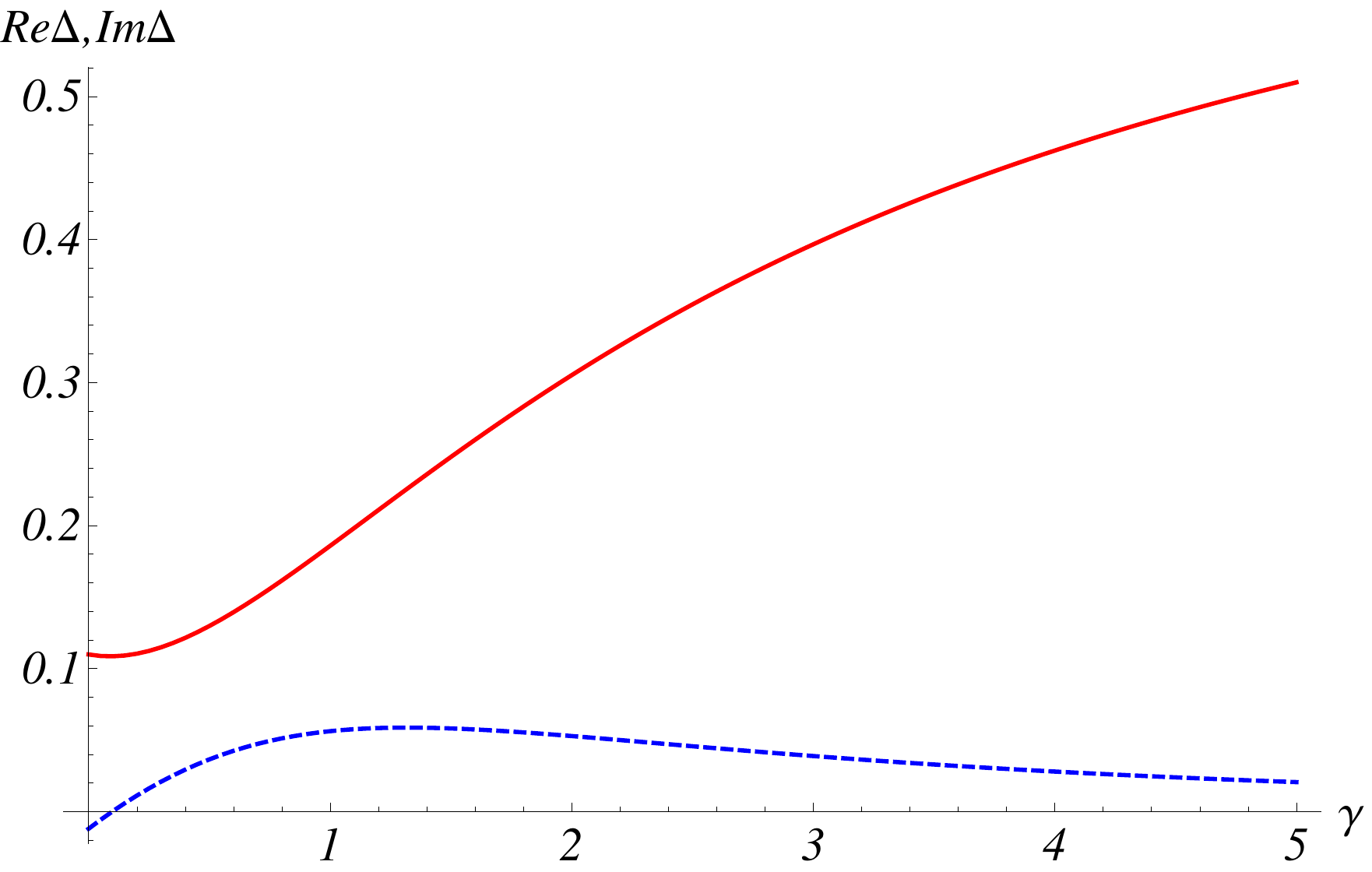}\\b}
\end{minipage}
\begin{minipage}[h]{0.5\linewidth}
\center{\includegraphics[width=1.0\linewidth]{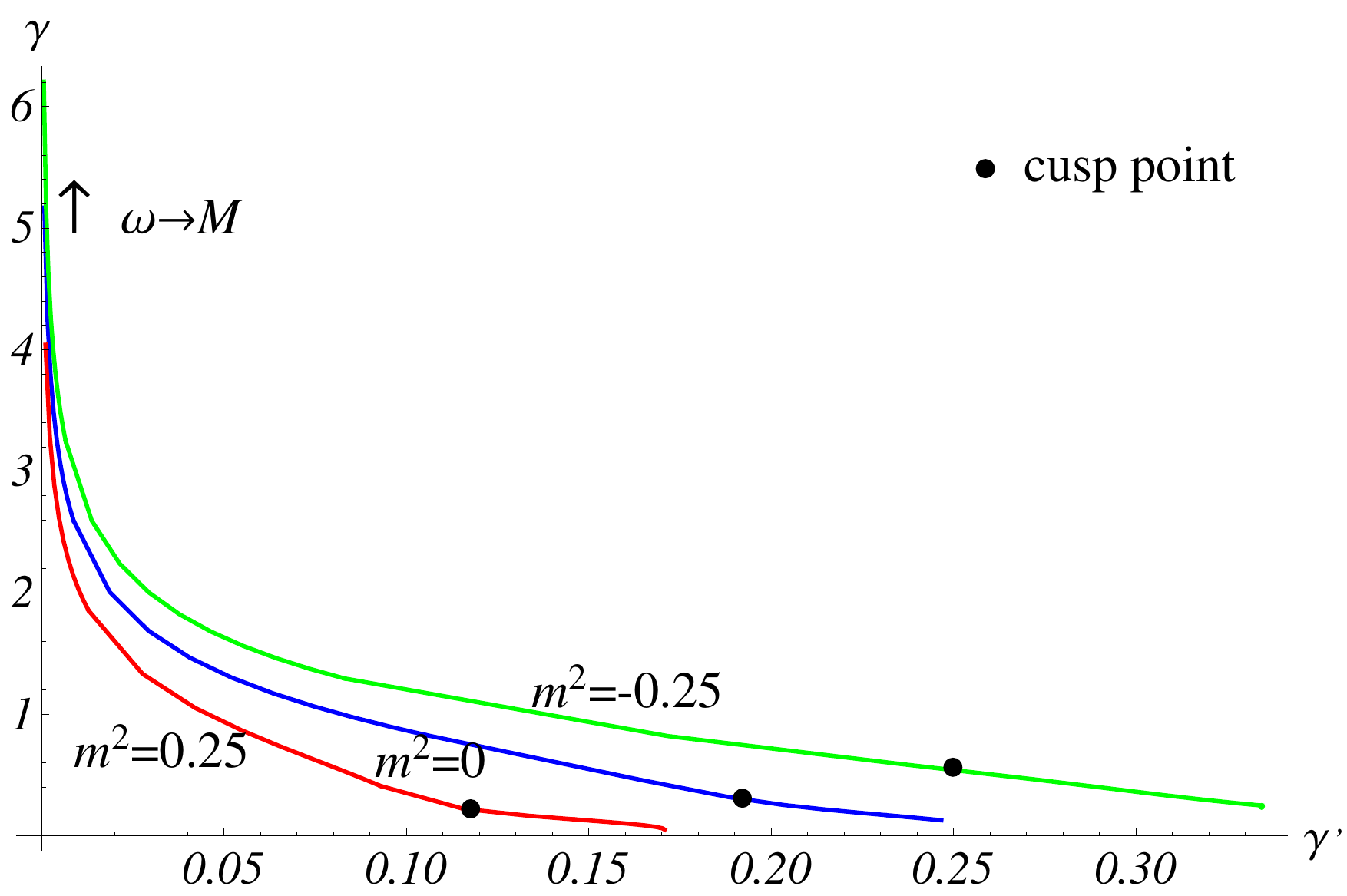}\\c}
\end{minipage}
\begin{minipage}[h]{0.5\linewidth}
\center{\includegraphics[width=1.0\linewidth]{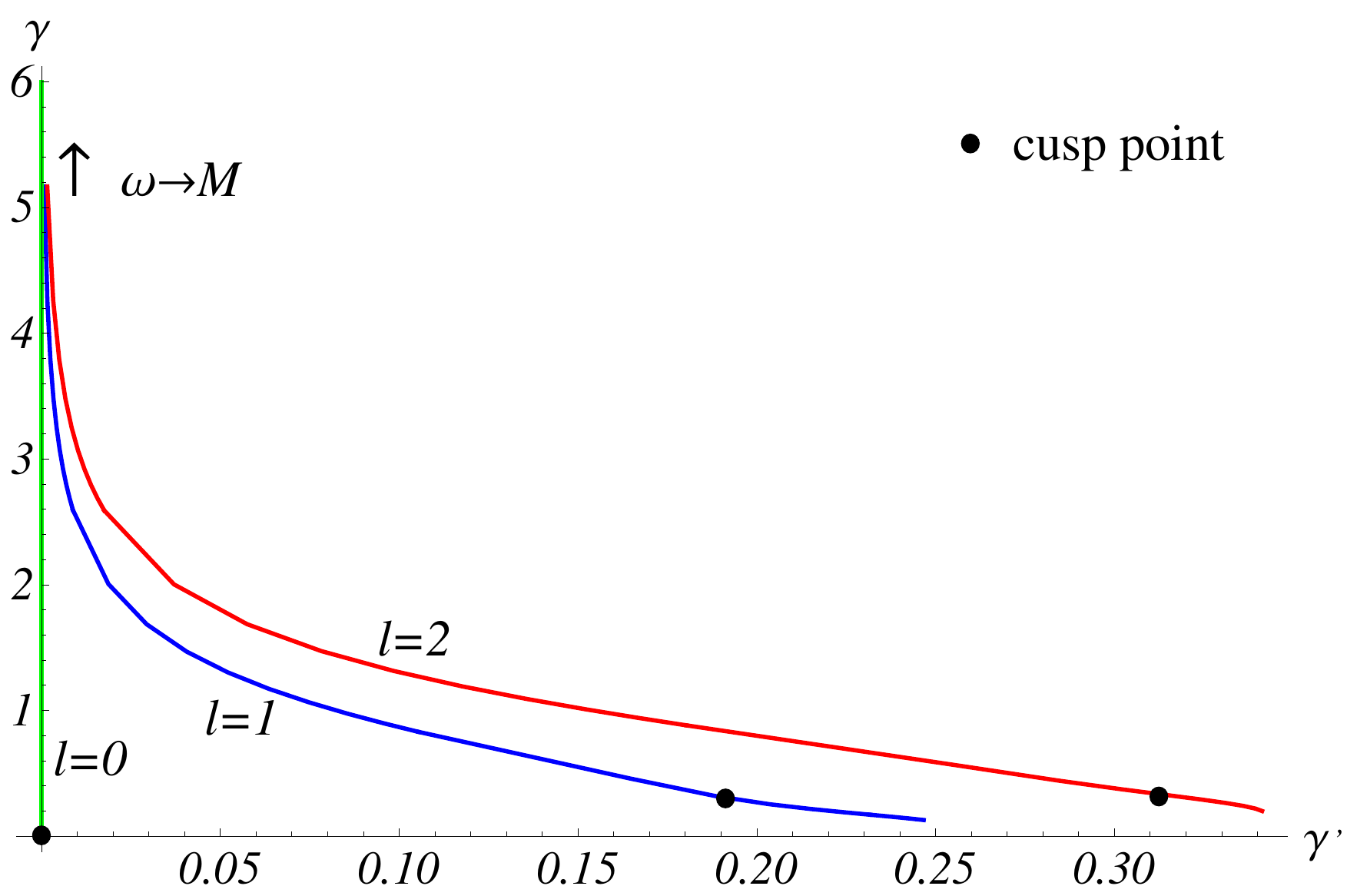}\\d}
\end{minipage}
\caption{a) The simultaneous root of the equations Re$\Delta(\gamma,\gamma')=0$ (solid line) and Im$\Delta(\gamma,\gamma')=0$ (dashed line) for $n=2$, $m=0$, $M=v=l=1$, $\omega=0.99$. Here $\gamma'\approx 0.07$. 
b) The absence of simultaneous roots for $n=0$.
c) The family of roots of the same equations for different values of $\omega$. Again, $n=2$, $l=1$. d) The family of roots of the same equations for different values of $l$. Here $n=2$ and $m=0$.}
\label{LastPlot}
\end{figure}

\subsection*{3.5 The case $\gamma'\neq 0$}

According to the form of the perturbation ansatz [Eq.(\ref{anzats})], the case 
$\gamma'\neq 0$
would correspond to the decay of Q-tube through the excited states.
Calculations show that the complex roots of the equation $\Delta(\gamma,\gamma')=0$ exist, at least, in a wide range of background Q-tube configurations. As an example, shown in the Fig.\ref{LastPlot}a is the root corresponding to a quite large value of $\omega$ and $l=1$. So, one can make sure about the instability of excited Q-tubes living on the upper branches of the $E(Q)$-plot. But it appears that even lower branches become unstable against the perturbations with $\gamma'\neq 0$. 
\begin{figure}[htb!]
\begin{minipage}[h]{0.5\linewidth}
\center{\includegraphics[width=1.0\linewidth]{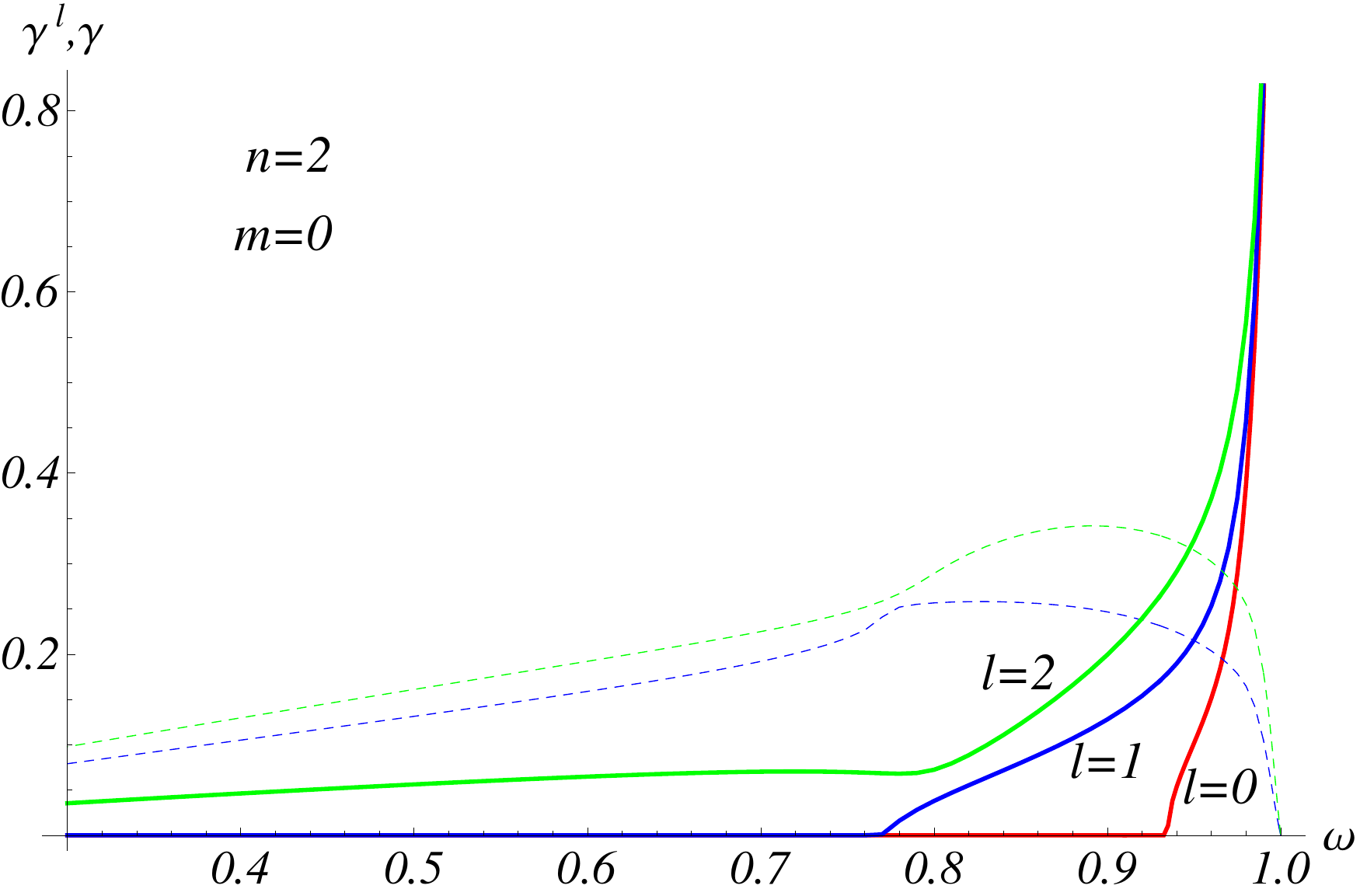}\\a}
\end{minipage}
\begin{minipage}[h]{0.5\linewidth}
\center{\includegraphics[width=1.0\linewidth]{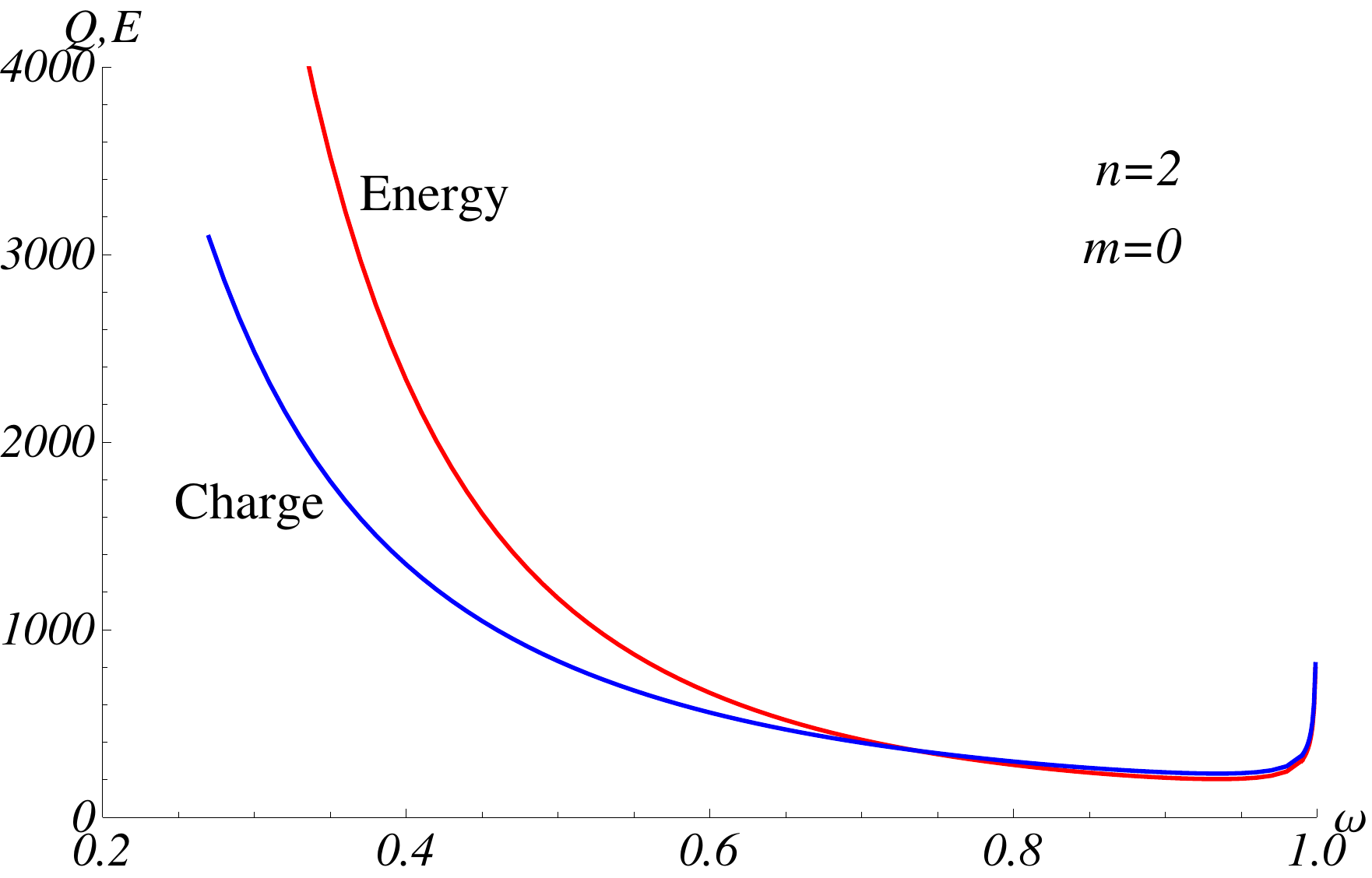}\\b}
\end{minipage}
\caption{ Scales of $\gamma'$ and $E$ for the considered values of $\omega$, with $n=2$ and $m=0$. a) $\gamma'$ (dashed) and $\gamma$ (solid) for modes with different $l$. b) Values of $E,Q$ in the same region of $\omega$. 
}
\label{scalesEgamma}
\end{figure}
To demonstrate this, we construct the parametric plots shown in Figs.\ref{LastPlot}c,d. Each point of these plots corresponds to the 
simultaneous roots at some value of $\omega$. We see that for a Q-tube with large $\omega$, it is sufficient to slightly excite it to obtain instability
with large value of $\gamma$. The lifetime $1/\gamma$ of the tube after such excitation tends to zero while $\omega\rightarrow M$. This agrees with what one can expect from upper-branch configurations. In contrast, the smaller the value of $\omega$, the larger the $\gamma'$ needed for the tube decay. We also check
that the typical values of $\gamma'\neq 0$ are small compared with the
energy density $E$ of the soliton. In Fig.\ref{scalesEgamma}, we show the dependence of 
$\gamma,\gamma',E$, and $Q$ on $\omega$. One can see that the energy scale of perturbations is suppressed to the scale of the solution by a factor of 
order $10^3$, and our classical consideration makes sense for this range of $\omega$.

We especially note that the foregoing is true for Q-tubes with $n\geqslant 1$. It turns out that for Q-ball-like configurations with $n=0$, one cannot obtain 
the simultaneous root for real and imaginary parts of $\Delta(\gamma,\gamma')$. This may signal the true stability of nonrotating tubes living on the lower branches of the $E(Q)$ plot; see
Fig.\ref{LastPlot}b.

\section*{CONCLUSION}

Our choice of potential allows us to construct a powerful quantitative method for the classical stability investigation. We survey instabilities in a wide range of parameters.
The criterion of the classical stability of Q-balls $\partial^2E/\partial Q^2<0$ remains
unchanged in the case $n=0$. However, using the explicit solution, we found a continuous family of instabilities for $n\geq 1$. 
Our result has an analogy with the superconductivity of the second type:
even if the solution is stable in the $l=0$ mode, the transition in
$l\geq 1$ mode is possible. We also check that the typical scale of the excitation
energy is suppressed by the soliton energy $E$, $\gamma'/E\sim 10^{-3}$,
and the classical analysis is appropriate for the entire range of considered $\omega$. 

\section*{ACKNOWLEDGMENTS}

The authors are indebted to M.~N.~Smolyakov  for reading the text and providing comments that improved the manuscript, and to M.~V.~Libanov for helpful discussions.
This work was supported by Grant No. NS-2835.2014.2 of the President
of the Russian Federation and by RFBR Grant No. 14-02-31384.

\bibliography{Tube_stab_4.04.14}
\bibliographystyle{unsrt}

\end{document}